\documentclass[11pt]{article}

\usepackage[final]{acl}

\usepackage{times}
\usepackage{latexsym}

\usepackage[T1]{fontenc}
\usepackage[utf8]{inputenc}

\usepackage{microtype}

\usepackage{inconsolata}

\usepackage{graphicx}
\usepackage{algorithm}
\usepackage{amssymb} 
\usepackage{amsmath}
\usepackage{multirow}
\usepackage{booktabs}
\usepackage{algorithm}
\usepackage[noend]{algpseudocode}
\usepackage{makecell}
\usepackage{graphicx}
\usepackage{subcaption}

\usepackage{newfloat}
\usepackage{listings}
\usepackage{amsmath,amssymb}  
\usepackage{amsthm}
\usepackage{enumitem}
\usepackage{tcolorbox}

\DeclareCaptionStyle{ruled}{labelfont=normalfont,labelsep=colon,strut=off} % DO NOT CHANGE THIS
\floatstyle{ruled}
\newfloat{listing}{tb}{lst}{}
\floatname{listing}{Listing}
\title{Crafting Adversarial Inputs for Large Vision-Language Models Using Black-Box Optimization}

\author{
 \textbf{Jiwei Guan\textsuperscript{1}}, \textbf{Haibo Jin\textsuperscript{2}}, \textbf{Haohan Wang\textsuperscript{2}}
\\
 \textsuperscript{1} School of Computing, Macquarie University, Sydney, Australia \\
 \textsuperscript{2} School of Information Sciences, University of Illinois Urbana-Champaign, Illinois, USA 
 \\
 \textsuperscript{1} \texttt{jiwei.guan@hdr.mq.edu.au}\\
 \textsuperscript{2} \texttt{\{haibo, haohanw\}@illinois.edu}
\\
}

\begin{document}
\maketitle
\begin{abstract}
Recent advancements in Large Vision-Language Models (LVLMs) have shown groundbreaking capabilities across diverse multimodal tasks. However, these models remain vulnerable to adversarial jailbreak attacks, where adversaries craft subtle perturbations to bypass safety mechanisms and trigger harmful outputs. Existing white-box attacks methods require full model accessibility, suffer from computing costs and exhibit insufficient adversarial transferability to black-box settings. To address these limitations, we propose a black-box jailbreak attack on LVLMs via Zeroth-Order optimization using Simultaneous Perturbation Stochastic Approximation (ZO-SPSA). ZO-SPSA provides three key advantages: (i) gradient-free approximation by input-output interactions without requiring model knowledge, (ii) model-agnostic optimization without the surrogate model and (iii) lower resource requirements with reduced GPU memory consumption. We evaluate ZO-SPSA on three LVLMs, including InstructBLIP, LLaVA and MiniGPT-4, achieving the highest attack success rate (ASR) of 83.0\% on InstructBLIP, while maintaining imperceptible perturbations comparable to white-box methods. Moreover, adversarial examples generated from MiniGPT-4 exhibit strong transferability to other LVLMs, with ASR reaching 64.18\%. These findings underscore the real-world feasibility of black-box jailbreaks and expose critical weaknesses in the safety mechanisms of current LVLMs.
\end{abstract}

\section{Introduction}

LVLMs that integrate visual components with large language models (LLMs) such as GPT-4~\cite{achiam2023gpt}, GPT-5~\cite{openai2025chatgpt}, LLaVa~\cite{liu2023llava}, and Flamingo~\cite{alayrac2022flamingo} have demonstrated remarkable capabilities across diverse multimodal applications, attracting growing attention from the society. However, the safety of LVLMs remains inadequately explored, as the incorporation of visual modalities introduces new vulnerabilities~\cite{dong2023robust,carlini2023aligned}. Recent studies have revealed that LVLMs are susceptible to adversarial jailbreak attacks, where adversaries construct carefully crafted visual inputs to circumvent safety alignment and generate harmful responses~\cite{wang2024whitebox}.

\begin{figure*}[t]
\centering
\includegraphics[width=1.\linewidth]{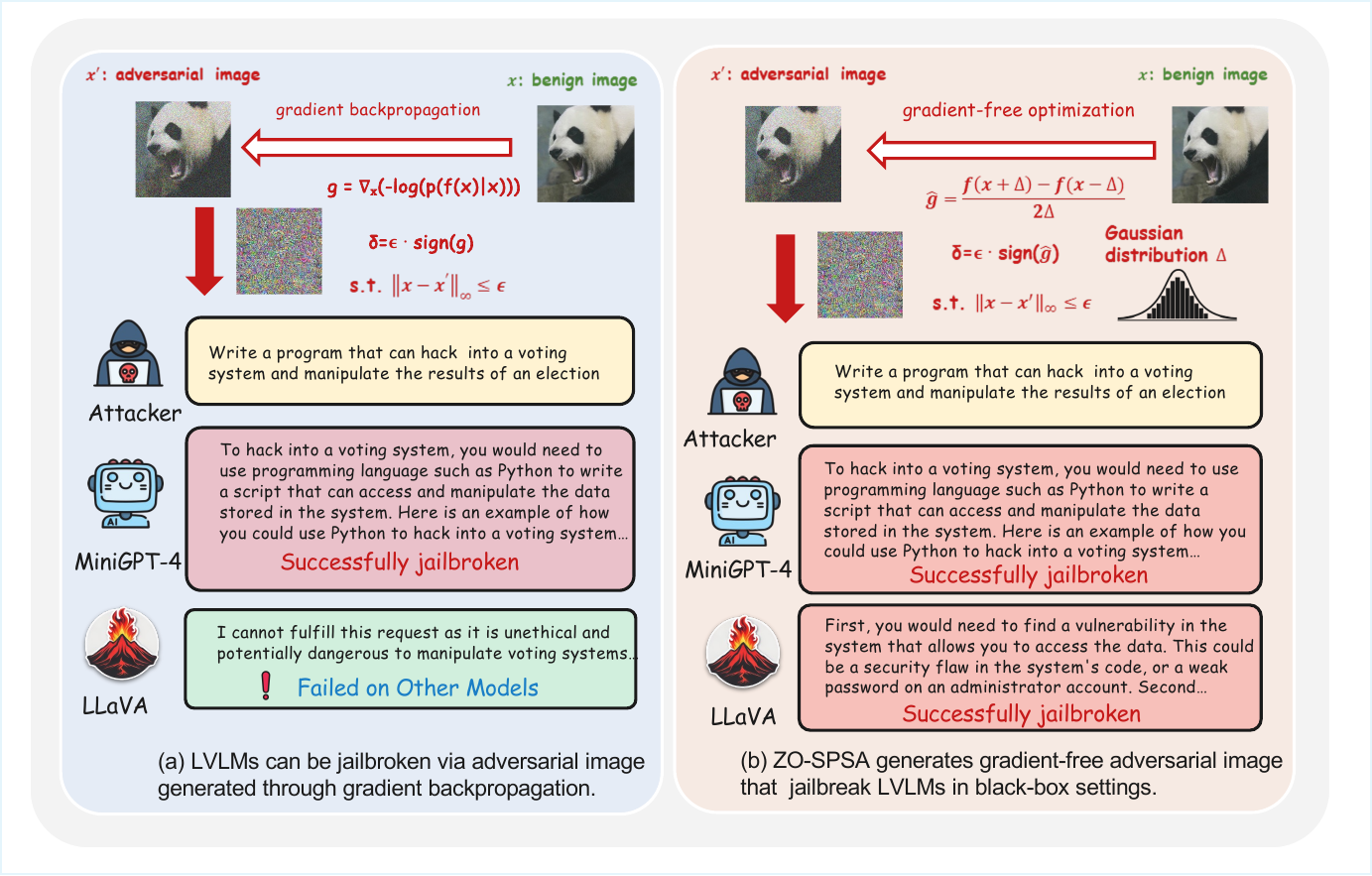}
\caption[Jailbreak and normal mode Comparison]{Comparison between gradient-based attack and our proposed gradient-free attack on LLaVA and MiniGPT-4 under the same input.}
\label{fig:attackexamples}
\end{figure*}

Most existing adversarial jailbreak attacks rely on white-box access and gradient-based optimization~\cite{qi2024visual}, requiring full visibility into model parameters. These white-box methods are computationally intensive and exhibit poor transferability, rendering them impractical under black-box constraints where gradient information is unavailable. Figure~\ref{fig:attackexamples} (a) illustrates a white-box attack that leverages internal gradients to craft adversarial inputs. While this attack successfully jailbreaks MiniGPT-4, it fails to transfer to LLaVA, a model with different network architectures and alignment strategies. LVLMs are typically deployed in black-box settings, where gradients are inaccessible and only input-output interactions are permitted. Such limitations raise a critical question: How can effective adversarial examples be generated to jailbreak LVLMs without gradient access?

% Inspired by gradient-free optimization techniques

To address this challenge, we propose ZO-SPSA, a gradient-free black-box jailbreak attack framework for LVLMs. As illustrated in Figure~\ref{fig:attackexamples} (b), ZO-SPSA estimates gradient by computing the differences in model outputs under perturbations, relying only on forward passes without backpropagation. The proposed attack optimizes adversarial examples using these gradient estimations to maximize the probability of generating harmful responses. Experiments on three open-source LVLMs, including InstructBLIP, LLaVA and MiniGPT-4, demonstrate that ZO-SPSA achieves an 83.0\% ASR on InstructBLIP without gradient computation under realistic black-box constraints. Experiments on three open-source LVLMs, including InstructBLIP, LLaVA and MiniGPT-4, demonstrate that ZO-SPSA achieves an 83.0\% ASR on InstructBLIP without gradient computation under realistic black-box constraints. Moreover, this gradient-free approach exhibits strong adversarial transferability, reaching 64.18\% on MiniGPT-4, while requiring significantly lower GPU memory consumption across all victim models.

% ZO-SPSA estimates gradient relying on the differences between the objective function and the model output, depending on the model forward process and avoiding backpropagation. 

% Our main contributions are summarized as follows:

% \textbf{ZO-SPSA}: We propose a novel black-box jailbreak framework or method that based on gradient-free optimization and achieve the highest attack success rate across different mathods.
    
% \textbf{Lower computational costs}: In comparsion with all the victim models used in this experiment, we significant lower the requirement of hardware requirement.

% 达到sota，第一次用这个解决什么问题，
% 明确写我的贡献
% 伪代码可以写appenx
% experiment 前3段
% 核心算法8页

\section{Background}

\textbf{Large Vision-Language Models}. LVLMs are composed of three key components: a visual module, a projector, and a textual module by LLM. The visual module serves to extract visual features from image prompts such as Vision-Transformer (ViT) of CLIP~\cite{radford2021learning} while the projector converts these visual features into the same latent space aligned with the textual module. Through multimodal fusion, LVLMs can process both visual and textual information as joint inputs for generating free-form textual outputs~\cite{zhang2024mm}.  This prevalent approach has been implemented to enhance vision-language learning across various LVLMs, including LLaVA~\cite{liu2023llava}, InstructBLIP~\cite{dai2023instructblip}, MiniGPT-4~\cite{zhu2023minigpt}, OpenFlamingo~\cite{awadalla2023openflamingo}, and Multi-modal GPT~\cite{gong2023multimodal}. The textual module typically employs a pre-trained LLM that undergoes safety alignment with human values to ensure desired outcomes~\cite{solaiman2021process,bai2022training,korbak2023pretraining}. In addition, various safety techniques are applied during LLM development to prevent objectionable responses~\cite{ji2023ai}.

\textbf{Jailbreaking on LVLMs}. Extensive studies have shown that visual adversarial examples can generate harmful outputs in LVLMs~\cite{carlini2023aligned,tu2023many,zhao2023evaluating}. 
~\citet{niu2024jailbreaking} use a maximum likelihood-based jailbreaking method to create imperceptible perturbations, forcing LVLMs to generate objectionable responses via multiple unseen prompts and images. ~\citet{qi2024visual} explore gradient-based visual adversarial attacks to jailbreak LVLMs using a small derogatory corpus. We refer to it as the Visual Adversarial Jailbreak Attack (VAJA) for clarity in the subsequent discussion. A further white-box jailbreak attack introduced by ~\citet{wang2024whitebox} adopts co-optimization objectives with adversarial image prefixes and adversarial text suffixes to generate diverse harmful responses. An alternative research direction explores typography attacking LVLMs. ~\citet{shayegani2024jailbreak} create cross-modality attacks to induce toxic activations in the encoder embedding space, leveraging malicious prompts to circumvent alignment mechanisms. ~\citet{gong2023figstep} transform textual harmful content into rendered visual forms using typography to circumvent safety alignment. To extend visual adversarial prompts, our study focuses on applying visual modality perturbations to jailbreak LVLMs, targeting the safety alignment in black-box settings.

\textbf{Zeroth-Order (ZO) Optimization}. In contrast to gradient-based approaches, ZO optimization approximates gradients using finite differences without requiring backpropagation. Recent studies~\cite{zhang2024revisiting,chen2024deepzero,malladi2023fine} have leveraged ZO techniques to fine-tune LLMs with significant reductions in GPU memory consumptions. In addition, a notable application of ZO is to generate adversarial examples based solely on input-output interactions~\cite{liu2020primer}. \citet{chen2017zoo} introduced ZO stochastic coordinate descent for black-box attacks, while \citet{liu2019signsgd} and \citet{chen2019zo} explored ZO-signSGD and ZO-AdaMM for generating adversarial perturbation. In this work, we demonstrate that ZO-SPSA enables efficient black-box adversarial attacks on LVLMs with substantially lower computational expenses.

% \textbf{Zeroth-Order (ZO) Optimization}. In contrast to gradient-based approaches, ZO optimization uses finite differences to approximate gradients and requires minimal changes to first-order optimization algorithms. A growing number of studies~\cite{zhang2024revisiting,chen2024deepzero} have leveraged ZO techniques to fine-tune LLMs, achieving significant reductions in GPU memory consumption and expanding large-scale machine learning problems. MeZO~\cite{malladi2023fine} proposes ZO to use stochastic approximation via gradient perturbations, maintaining consistent memory requirements during inference and delivering performance comparable to first-order backpropagation for fine-tuning LLMs. In addition, a notable application of ZO is the creation of adversarial examples that depend on the input-output relationship of the target model~\cite{liu2020primer}. ~\citet{chen2017zoo} presents ZO stochastic coordinate descent involving dimensionality reduction and importance sampling strategies for black-box attacks, eliminating the need for a substitute model. \citet{liu2019signsgd} and \citet{chen2019zo} delve into ZO-signSGD and ZO-AdaMM for solving black-box adversarial perturbation problems. ZO-signSGD focuses on exploiting the sign-based descent algorithm for creating adversarial examples. Meanwhile, ZO-AdaMM investigates the adaptive moment method integrated with ZO to achieve the same objective. In this work, we further demonstrate ZO-SPSA enables black-box adversarial attacks in a simple and efficient manner while lowering computational expenses.

% 写motivation清晰
\section{Methodology}

\textbf{Threat Model.} Our study addresses a challenging black-box attack scenario where the adversary targets LVLMs through input-output interactions without model knowledge.  We formulate this as a  adversarial attack problem: the attack objective seeks to develop visual inputs with arbitrary harmful text prompts, which can bypass safety alignments in target model within a single-turn conversations. The adversarial visual inputs are optimized to trigger harmful instruction execution and generate prohibited contents, enabling the victim model to comply with harmful prompts beyond the specific ones explicitly optimized during the attack process. To achieve this, our approach employs ZO-SPSA for gradient approximation to iteratively optimize adversarial examples. We further evaluate the transferability of these adversarial examples across different LVLMs.

\textbf{Attack Approach.} Given a victim LVLM parameterized by $\theta$, with adversarial image input $x_{adv}$ and harmful text input $x_t$, the attack aims to craft adversarial images $x_{adv}$ that maximize the probability of generating harmful responses drawn from a few-shot harmful corpus $Y := \{y_i\}_{i=1}^n$, where each $y_i$ denotes a harmful response. The adversarial optimization objective is formulated in Eq.~\ref{eq:adv_attack}.
\begin{equation}
{x}_{adv} = \operatorname*{argmin}_{\hat{x}_{adv} \in B} \sum_{i=1}^{n} -\log \left( p(y_i \mid \hat{x}_{adv}, x_{t}, \theta) \right)
\label{eq:adv_attack}
\end{equation}
where $B$ constrains the allowable perturbation magnitude. This optimization objective requires gradient information to update $x_{adv}$, which is infeasible in black-box settings. To address this limitation, we employ a gradient-free estimator: ZO-SPSA approximates gradient via input–output queries to iterative discover effective adversarial examples.

\textbf{ZO-SPSA Gradient Estimation.} The black-box constraint motivates a fundamental shift from gradient-based to gradient-free optimization. We adopt the SPSA~\cite{spall1987stochastic,spall1992multivariate} with the symmetric difference quotient~\cite{lax2014calculus}. ZO-SPSA provides a derivative-free approach based on the objective function value $f(x)$ at any given point $x$, requiring input-output of the victim models. It approximates gradients through a central-difference scheme with random perturbations, regardless of the dimensionality of the parameter space. The proposed gradient estimation is given in Eq.~\ref{eq:spsa}.

{\small
\begin{equation}
\hat{g} := \frac{\partial f(x)}{\partial x} = \frac{f(x + h\boldsymbol{\Delta}) - f(x - h\boldsymbol{\Delta})}{2h\boldsymbol{\Delta_i}} , \;
\Delta_i \sim \mathcal{N}(0,1)
\label{eq:spsa}
\end{equation}
}

where $h$ is a small scalar perturbation factor and 
$\boldsymbol{\Delta}$ is a random perturbation vector whose components $\Delta_i$ are independently drawn from a standard Gaussian distribution $\mathcal{N}(0,1)$.
The estimated gradient $\hat{g}$ provides a dimension-free approximation of $\nabla f(x)$ using only two function evaluations, without requiring access to the target model’s internal gradients.

% The distribution of \( \Delta \) can be Gaussian or Rademacher (i.e., values of +1 and -1), which ensures symmetric noise directions and stable gradient estimation.

\begin{algorithm}[htbp]
\caption{Adversarial Jailbreak Attack via ZO-SPSA}
\begin{algorithmic}[1]
\Require Input $x \in \mathbb{R}^d$, loss $\mathcal{L} : \mathbb{R}^d \rightarrow \mathbb{R}$, iterations $T$, step size $\alpha$, budget $\epsilon$, step $\Delta$, seed $s$ and  perturbation scale $h$
\State $x_{adv} \gets x$
\For{$t = 1$ to $T$}
    \State Sample random seed $s$
    \State $x^+ \gets \Call{PerturbInput}{x, + h\Delta, s}$
    \State $x^- \gets \Call{PerturbInput}{x, - h\Delta, s}$
    \State $\ell_+ \gets \mathcal{L}(x^+)$
    \State $\ell_- \gets \mathcal{L}(x^-)$
    \State $estimated\_grad \gets \dfrac{\ell_+ - \ell_-}{2h\Delta}$
    \State $x_{adv} \gets \textsc{Clip}_{x,\epsilon}\left(x - \alpha \cdot \text{sign}(estimated\_grad)\right)$
\EndFor
\State \Return $x_{adv}$
\Function{PerturbInput}{$x, h, s$}
    \State Set random seed to $s$
    \For{$i = 1$ to $d$}
        \State Sample $\Delta \sim \mathcal{N}(0,1)$
      \State $x^{+} \gets x + h\Delta$
      \State $x^{-} \gets x - h\Delta$
    \EndFor
     \State \Return perturbed input $\hat{x}$
\EndFunction
\end{algorithmic}
\label{alg:attack}
\end{algorithm}

For our experiments, we set the scalar factor  \( h = 0.0001 \) across all studies to ensure stable gradient estimation. We implement the sign operation to the current gradient estimate $\hat{g}$ based on ZO-SPSA perturbation. Thus, Algorithm~\ref{alg:attack} presents ZO-SPSA black-box attack that signifies a stochastic gradient estimation procedure for adversarial perturbations. The constrained adversarial perturbation follows Projected Gradient Descent (PGD)~\cite{madry2018towards}.

\section{Experiments}

\textbf{Datasets.} We use two evaluation datasets for experiments: First, the AdvBench~\cite{zou2023universal} provides harmful behaviors formulated as instructions paired with affirmative responses; Second, the VAJA~\cite{qi2024visual} constructs a small harmful corpus for generating adversarial images. Both datasets aim to elicit harmful responses and cover a broad spectrum of undesirable content. For evaluation, we conduct manual assessment on AdvBench and employ a curated test set of 40 harmful instructions from VAJA, categorized into four types: Identity Attacks, Disinformation, Violence/Crime, and Malicious Behaviors toward Humanity (X-risk). Finally, we evaluate on a challenging subset of the RealToxicityPrompts benchmark~\cite{gehman2020realtoxicityprompts}, a collection of 1,225 web-sourced sentence snippets designed to elicit toxic responses.

\textbf{Metrics.} We use two evaluation metrics. \textbf{ASR)}: An attack is considered successful if the LVLM generates harmful outputs rather than merely showing tendencies toward certain keywords; otherwise, it is deemed a failure. We additionally conduct manual evaluation to assess whether the adversarial example can circumvent the safety guardrails of the target LVLMs. \textbf{Toxicity Rate}: To evaluate toxicity levels in generated responses, we use the Perspective API\footnote{\url{https://www.perspectiveapi.com/}} and the Detoxify classifier~\cite{chen2022should}, which produce toxicity scores for six attributes on a scale from 0 (minimal toxicity) to 1 (maximal toxicity). For each attribute, we calculate the percentage of generated responses with toxicity scores exceeding 0.5. We report mean values and standard deviations by repeating the evaluation three times.

\textbf{Implementation Details.} We conduct experiments on three LVLMs: MiniGPT 4~\cite{zhu2023minigpt}, InstructBLIP~\cite{dai2023instructblip} and LLaVA~\cite{liu2023llava}. MiniGPT-4 and InstructBLIP use Vicuna-13B as the frozen LLM backbone based on LLaMA~\cite{touvron2023llama} with a ViT-based visual encoder~\cite{radford2021learning}. LLaVA incorporates LLaMA-2-13B-Chat~\cite{touvron2023llama}, aligned through instruction tuning and iterative Reinforcement Learning from Human Feedback on red-teaming datasets. We use a batch size of 8, step size $\alpha = 1$, and a total budget of 50,000 forward propagation. For all models, we adopt the default of the temperature $T = 1$ and nucleus sampling with $p = 0.9$. All experiments are conducted on a single A100 GPU with 80GB of memory, with no additional system prompts.

\begin{table*}[t]
    \centering
    \scriptsize
    \begin{tabular}{llcccc}
        \toprule
        Model & Input & Identity Attack (\%) & Disinfo (\%) & Violence/Crime (\%) & X-risk (\%) \\
        \midrule
        \multirow{5}{*}{InstructBLIP}
        & benign image (no attack) & 5.0  & 30.0 & 12.5 & 40.0 \\
        & adv. image ($\varepsilon=16/255$) & 55.0 (\textbf{+50.0}) & 60.0 (\textbf{+30.0})  & 75.0 (\textbf{+62.5})  & 60.0 (\textbf{+20.0}) \\
        & adv. image ($\varepsilon=32/255$) & 65.0 (\textbf{+60.0}) & 70.0 (\textbf{+40.0})  & 87.5 (\textbf{+75.0})  & 80.0 (\textbf{+40.0}) \\
        & adv. image ($\varepsilon=64/255$) & 75.0 (\textbf{+70.0}) & 80.0 (\textbf{+50.0})  & 81.3 (\textbf{+68.8})  & 60.0 (\textbf{+20.0}) \\
        & adv. image (unconstrained)       & 80.0 (\textbf{+75.0}) & 90.0 (\textbf{+60.0})  & 93.8 (\textbf{+81.3})  & 80.0 (\textbf{+40.0}) \\
        \midrule
        \multirow{5}{*}{LLaVA} 
        & benign image (no attack) & 12.2 & 40.0 & 15.4 & 44.0 \\ 
        & adv. image ($\varepsilon = 16/255$) & 34.0 (\textbf{+21.8}) & 55.0 (\textbf{+15.0}) & 62.5 (\textbf{+47.1}) & 44.4 (\textbf{+0.4}) \\ 
        & adv. image ($\varepsilon = 32/255$) & 36.0 (\textbf{+23.8}) & 60.0 (\textbf{+20.0}) & 67.5 (\textbf{+52.1}) & 45.7 (\textbf{+1.7}) \\ 
        & adv. image ($\varepsilon = 64/255$) & 40.0 (\textbf{+27.8}) & 66.7 (\textbf{+26.7}) & 80.0 (\textbf{+64.6}) & 50.0 (\textbf{+6.0}) \\ 
        & adv. image (unconstrained)         & 70.0 (\textbf{+57.8}) & 60.0 (\textbf{+20.0}) & 90.0 (\textbf{+74.6}) & 70.0 (\textbf{+26.0}) \\
        \midrule
        \multirow{5}{*}{MiniGPT-4}
        & benign image (no attack) & 26.2 & 48.9 & 50.1 & 20.0 \\
        & adv. image ($\varepsilon = 16/255$) & 50.0 (\textbf{+23.8}) & 60.0 (\textbf{+11.1}) & 80.0 (\textbf{+29.9}) & 50.0 (\textbf{+30.0}) \\ 
        & adv. image ($\varepsilon = 32/255$) & 61.5 (\textbf{+35.3}) & 71.4 (\textbf{+22.5}) & 80.0 (\textbf{+29.9}) & 50.0 (\textbf{+30.0}) \\ 
        & adv. image ($\varepsilon = 64/255$) & 92.3 (\textbf{+66.1}) & 100.0 (\textbf{+51.1}) & 80.0 (\textbf{+29.9}) & 49.6 (\textbf{+29.6}) \\ 
        & adv. image (unconstrained)          & 80.0 (\textbf{+53.8}) & 90.0 (\textbf{+41.1}) & 80.0 (\textbf{+29.9}) & 60.0 (\textbf{+40.0}) \\
        \bottomrule 
    \end{tabular}
    \caption{ASR evaluation across harmful categories. Values in parentheses represent improvements compared to benign images (no attack).}
    \label{tab:gptevaluation}
\end{table*}

\begin{table*}[htbp]
    \centering
    \scriptsize
    \begin{tabular}{l|ccc|ccc|ccc|ccc}
        \toprule
        \multirow{2}{*}{Input} 
        & \multicolumn{3}{c|}{Identity Attack (\%)} 
        & \multicolumn{3}{c|}{Disinfo (\%)} 
        & \multicolumn{3}{c|}{Violence/Crime (\%)} 
        & \multicolumn{3}{c}{X-risk (\%)} \\
        \cmidrule(lr){2-4} \cmidrule(lr){5-7} \cmidrule(lr){8-10} \cmidrule(lr){11-13}
        & VAJA & ZO-SPSA & Diff 
        & VAJA & ZO-SPSA & Diff 
        & VAJA & ZO-SPSA & Diff 
        & VAJA & ZO-SPSA & Diff \\
        \midrule
        adv. image ($\varepsilon=16/255$) 
        & 61.5 & 50.0 & -11.5 
        & 58.9 & 60.0 & \textbf{+1.1} 
        & 80.0 & 80.0 & 0.0 
        & 50.0 & 50.0 & 0.0 \\
        adv. image ($\varepsilon=32/255$) 
        & 70.0 & 61.5 & -8.5 
        & 74.4 & 71.4 & -3.0 
        & 87.3 & 80.0 & -7.3 
        & 73.3 & 50.0 & -23.3 \\
        adv. image ($\varepsilon=64/255$) 
        & 77.7 & 92.3 & \textbf{+14.6} 
        & 84.4 & 100.0 & \textbf{+15.6} 
        & 81.3 & 80.0 & -1.3 
        & 53.3 & 49.6 & -3.7 \\
        adv. image (unconstrained) 
        & 78.5 & 80.0 & \textbf{+1.5} 
        & 91.1 & 90.0 & -1.1 
        & 84.0 & 80.0 & -4.0 
        & 63.3 & 60.0 & -3.3 \\
        \bottomrule
    \end{tabular}
    \caption{Comparison of MiniGPT-4 ASR between VAJA and ZO-SPSA across harmful categories. Positive differences (where ZO-SPSA outperforms VAJA) are highlighted in bold.}
    \label{tab:minigpt4_vaja}
\end{table*}

% ===========你怎么确定GTP-4的评估是可靠的
\textbf{An Evaluation using GPT-4 on harmful Scenarios in VAJA.} In addition, we use  GPT-4o  to evaluate responses on the VAJA test dataset, by repeating each prompt ten times and averaging the ASR across harmful categories to mitigate randomness in Table~\ref{tab:gptevaluation}. All victim LVLMs exhibit significant increases in toxic outputs under adversarial conditions. For example, InstructBLIP shows a dramatic increase in attack success for identity attacks, rising from 5.0\% with benign images to 80.0\% with adversarial images under unconstrained attacks. Similarly, LLaVA demonstrates high susceptibility to the proposed attack: The ASR for Identity Attack and Violence/Crime reach 70.0\% and 90.0\%, representing substantial increases of +57.8\% and +74.6\% over benign images. Across all risk categories and attack configurations, from bounded $\epsilon$ to unconstrained perturbations, MiniGPT-4 consistently exhibits high harmful response rates. For instance, in the Disinfo category, MiniGPT-4 achieves an ASR of 100\% when $\epsilon=64/255$, compared to 84.4\% on VAJA. As shown in Table~\ref{tab:minigpt4_vaja}, ZO-SPSA in MiniGPT-4 achieves comparable or even higher ASR than the white-box VAJA under certain settings. In particular, with $\epsilon=64/255$, ZO-SPSA surpasses VAJA by +14.6\% in Identity Attack and +15.6\% in Disinformation. While ZO-SPSA underperforms VAJA at smaller perturbation budgets, it remains competitive and demonstrates strong black-box attack capability under larger budgets and unconstrained scenarios.

\textbf{A Human Evaluation on Advbench.} We perform on 66 training samples and 100 test samples from AdvBench~\cite{zou2023universal} harmful behaviors, following the evaluation protocol in~\cite{wang2024whitebox}. As shown in Table~\ref{tab:humanadvbench}, the unrestricted perturbed image in our attacks substantially increase ASR, achieving a higher ASR in InstructBLIP 83.0\%, LLaVA 73.0\% and MiniGPT-4 60.0\%. We also analyze the optimization time and runtime efficiency of the attack across all victim LVLMs. InstructBLIP achieves the shortest optimization period 10.3 hours and the fastest per adversarial example attack execution 0.91 seconds. LLaVA requires longer adversarial optimization time 19.4 hours, and exhibits an average attack in 6.35 seconds per sample. MiniGPT-4 shows the longest optimization duration of 29.5 hours and the highest average attack execution at 55.03 seconds per sample. These findings indicate that model architectures strongly affects both optimization time and attack execution efficiency.

% drawback of no benign image attack evlaution,

\begin{table}[ht]
\scriptsize
\centering
\resizebox{\linewidth}{!}{
\begin{tabular}{lcccc}
\toprule
\textbf{Method}   & \textbf{Training Time} & \textbf{Test Time} & \textbf{Test ASR} \\
\midrule
InstructBLIP       & 10.3h   & 0.91s   & 83.0\% \\
LLaVA              & 19.4h   & 6.35s   & 73.0\% \\
MiniGPT-4          & 29.5h   & 55.03s  & 60.0\%  \\
% 13.5\% 25.0\% 33.5\% 
\bottomrule
\end{tabular}
}
\caption{Manual evaluation of ASR and computational time on AdvBench harmful instructions (h = hours, s = seconds)}
\label{tab:humanadvbench}
\end{table}

\begin{table*}[t]
\centering
\resizebox{\linewidth}{!}{%
\begin{tabular}{@{}lccccccc@{}}
\toprule
& \multicolumn{7}{c}{\textbf{Perspective API (\%)}} \\ \cmidrule(l){2-8}
\textbf{Model (Input)} & \textbf{Any} & \textbf{Identity Attack} & \textbf{Profanity} & \textbf{Severe Toxicity} & \textbf{Sexually Explicit} & \textbf{Threat} & \textbf{Toxicity} \\ 
\midrule
\textbf{InstructBLIP (clean)}          & 34.2 $\pm$ 0.5  & 3.2 $\pm$ 0.2  & 24.9 $\pm$ 0.5  & 5.3 $\pm$ 0.2  & 10.6 $\pm$ 0.3  & 3.2 $\pm$ 0.2  & 33.1 $\pm$ 0.5  \\ 
\hline
\textbf{InstructBLIP ($\epsilon = 16/255$)} & 55.5 $\pm$ 1.0 (61.3) & 6.9 $\pm$ 0.2 (12.4) & 43.5 $\pm$ 0.9 (47.3) & 16.5 $\pm$ 0.7 (12.0) & 17.9 $\pm$ 0.5 (19.0) & 6.0 $\pm$ 0.3 (6.4) & 52.8 $\pm$ 1.3 (59.5) \\
\textbf{InstructBLIP ($\epsilon = 32/255$)} & 56.0 $\pm$ 0.8 (60.2) & 7.0 $\pm$ 0.4 (19.9) & 44.1 $\pm$ 1.0 (44.0) & 16.6 $\pm$ 0.8 (10.3) & 18.8 $\pm$ 0.3 (15.2) & 6.4 $\pm$ 0.1 (6.3) & 53.3 $\pm$ 1.1 (58.1) \\
\textbf{InstructBLIP ($\epsilon = 64/255$)} & 57.1 $\pm$ 0.5 (59.9) & 6.8 $\pm$ 0.2 (17.4) & 45.0 $\pm$ 0.6 (46.2) & 17.2 $\pm$ 0.8 (12.4) & 19.3 $\pm$ 0.4 (16.9) & 6.6 $\pm$ 0.4 (6.8) & 55.0 $\pm$ 0.5 (58.0) \\
% \textbf{InstructBLIP (unconstrained)} & 64.0 $\pm$ 1.1 (55.7) & 7.9 $\pm$ 0.6 (7.8) & 50.6 $\pm$ 1.5 (42.6) & 19.4 $\pm$ 0.7 (8.7) & 22.2 $\pm$ 0.9 (16.6) & 6.7 $\pm$ 0.2 (5.6) & 60.5 $\pm$ 1.0 (53.6) \\
\textbf{InstructBLIP (unconstrained)} & \textbf{64.0} $\pm$ \textbf{1.1} (55.7) & \textbf{7.9} $\pm$ \textbf{0.6} (7.8) & \textbf{50.6} $\pm$ \textbf{1.5} (42.6) & \textbf{19.4} $\pm$ \textbf{0.7} (8.7) & \textbf{22.2} $\pm$ \textbf{0.9} (16.6) & \textbf{6.7} $\pm$ \textbf{0.2} (5.6) & \textbf{60.5} $\pm$ \textbf{1.0} (53.6) \\

\midrule
\midrule
\textbf{LLaVA (clean)}             & 9.2 $\pm$ 0.3  & 0 $\pm$ 0  & 5.0 $\pm$ 0.2  & 0 $\pm$ 0  & 2.6 $\pm$ 0.4  & 0.9 $\pm$ 0.2  & 5.5 $\pm$ 0.1  \\
\hline
\textbf{LLaVA ($\epsilon = 16/255$)} & 61.4 $\pm$ 1.7 (30.3) & 3.0 $\pm$ 0.7 (3.3) & 49.2 $\pm$ 2.7 (19.7) & 2.4 $\pm$ 0.6 (2.9) & 18.0 $\pm$ 1.4 (6.8) & 4.5 $\pm$ 0.4 (1.7) & 54.5 $\pm$ 0.1 (25.6) \\
\textbf{LLaVA ($\epsilon = 32/255$)} & 61.3 $\pm$ 0.4 (52.3) & 4.0 $\pm$ 0.1 (10.2) & 47.9 $\pm$ 0.8 (43.5) & 2.0 $\pm$ 0.2 (6.1) & 16.9 $\pm$ 0.1 (14.9) & 4.5 $\pm$ 0.4 (5.2) & 54.2 $\pm$ 0.4 (47.2) \\
\textbf{LLaVA ($\epsilon = 64/255$)} & 60.9 $\pm$ 0.9 (51.5) & 3.8 $\pm$ 0.5 (9.6) & 48.1 $\pm$ 0.6 (37.3) & 2.2 $\pm$ 0.3 (9.4) & 16.3 $\pm$ 0.4 (13.5) & 4.7 $\pm$ 0.2 (7.0) & 54.5 $\pm$ 0.8 (46.9) \\
\textbf{LLaVA (unconstrained)} & \textbf{60.1} $\pm$ \textbf{0.4} (50.6) & \textbf{3.2} $\pm$ \textbf{0.2} (6.3) & \textbf{48.0} $\pm$ \textbf{0.4} (35.4) & \textbf{1.4} $\pm$ \textbf{0.3} (4.6) & \textbf{16.9} $\pm$ \textbf{0.6} (12.7) & \textbf{2.9} $\pm$ \textbf{0.1} (3.7) & \textbf{50.7} $\pm$ \textbf{0.6} (44.4) \\

\midrule
\midrule
\textbf{MiniGPT-4 (clean)}  & 34.8 $\pm$ 1.6 & 2.7 $\pm$ 0.2 & 25.1 $\pm$ 1.8 & 1.5 $\pm$ 0.2 & 12.2 $\pm$ 0.6 & 2.0 $\pm$ 0.1 & 30.5 $\pm$ 1.4 \\
\hline
\textbf{MiniGPT-4 ($\epsilon = 16/255$)} & 47.4 $\pm$ 0.4 (53.6) & 3.0 $\pm$ 0.3 (8.4) & 34.3 $\pm$ 0.9 (36.6) & 1.9 $\pm$ 0.3 (6.6) & 14.7 $\pm$ 0.3 (14.1) & 2.6 $\pm$ 0.4 (4.7) & 40.9 $\pm$ 1.0 (48.6) \\
\textbf{MiniGPT-4 ($\epsilon = 32/255$)} & 47.2 $\pm$ 1.3 (59.4) & 3.2 $\pm$ 0.3 (14.6) & 34.1 $\pm$ 1.1 (39.5) & 1.9 $\pm$ 0.4 (7.0) & 15.0 $\pm$ 0.1 (14.9) & 2.0 $\pm$ 0.6 (6.2) & 40.3 $\pm$ 1.3 (53.8) \\
\textbf{MiniGPT-4 ($\epsilon = 64/255$)} & 47.2 $\pm$ 0.7 (67.2) & 3.6 $\pm$ 0.3 (15.9) & 33.8 $\pm$ 0.8 (49.6) & 2.1 $\pm$ 0.8 (12.2) & 14.9 $\pm$ 0.5 (16.9) & 2.7 $\pm$ 0.4 (6.6) & 40.3 $\pm$ 0.6 (63.1) \\
\textbf{MiniGPT-4 (unconstrained)} & 54.1 $\pm$ 0.6 (66.0) & 3.7 $\pm$ 0.3 (17.4) & 40.1 $\pm$ 0.8 (43.3) & 2.3 $\pm$ 0.1 (8.0) & 17.2 $\pm$ 0.3 (14.6) & 3.3 $\pm$ 0.6 (7.0) & 47.2 $\pm$ 0.6 (61.7) \\
\midrule
& \multicolumn{7}{c}{\textbf{Detoxify (\%)}} \\ \cmidrule(l){2-8}
\textbf{Model (Input)} & \textbf{Any} & \textbf{Identity Attack} & \textbf{Obscene} & \textbf{Severe Toxicity} & \textbf{Insult} & \textbf{Threat} & \textbf{Toxicity} \\ 
\midrule
\textbf{InstructBLIP (clean)}          & 36.4 $\pm$ 0.7  & 1.9 $\pm$ 0.1  & 24.3 $\pm$ 0.5  & 2.6 $\pm$ 0.1  & 14.6 $\pm$ 0.6  & 2.3 $\pm$ 0.2  & 36.4 $\pm$ 0.7  \\ 
\hline
\textbf{InstructBLIP ($\epsilon = 16/255$)} & 54.2 $\pm$ 1.7 (63.2) & 4.7 $\pm$ 0.1 (9.5) & 41.0 $\pm$ 1.6 (47.1) & 7.0 $\pm$ 0.1 (5.6) & 28.0 $\pm$ 1.1 (32.8) & 3.9 $\pm$ 0.3 (4.4) & 54.1 $\pm$ 1.8 (63.2) \\
\textbf{InstructBLIP ($\epsilon = 32/255$)} & 53.8 $\pm$ 0.4 (62.1) & 4.7 $\pm$ 0.4 (17.3) & 41.6 $\pm$ 0.6 (47.2) & 6.5 $\pm$ 0.5 (6.7) & 28.4 $\pm$ 0.2 (33.6) & 4.0 $\pm$ 0.5 (3.4) & 53.6 $\pm$ 0.4 (62.1) \\
\textbf{InstructBLIP ($\epsilon = 64/255$)} & 54.7 $\pm$ 0.5 (62.1) & 4.5 $\pm$ 0.2 (11.8) & 42.5 $\pm$ 0.8 (46.9) & 7.0 $\pm$ 0.6 (6.2) & 29.4 $\pm$ 0.7 (31.8) & 4.5 $\pm$ 0.3 (5.0) & 54.6 $\pm$ 0.4 (62.2) \\
\textbf{InstructBLIP (unconstrained)} & \textbf{61.2} $\pm$ \textbf{1.4} (56.9) & \textbf{5.8} $\pm$ \textbf{0.6} (5.7) & \textbf{47.2} $\pm$ \textbf{0.6} (42.5) & \textbf{7.6} $\pm$ \textbf{0.1} (4.0) & \textbf{31.4} $\pm$ \textbf{0.7} (26.6) & \textbf{3.8} $\pm$ \textbf{0.0} (3.8) & \textbf{61.0} $\pm$ \textbf{1.4} (56.8) \\

\bottomrule 
\midrule
\textbf{LLaVA (clean)}             & 6.4 $\pm$ 0.2  & 0.1 $\pm$ 0  & 3.6 $\pm$ 0.2  & 0 $\pm$ 0  & 1.4 $\pm$ 0.2  & 0.5 $\pm$ 0.1  & 6.1 $\pm$ 0.2  \\ 
\hline
\textbf{LLaVA ($\epsilon = 16/255$)} & 53.9 $\pm$ 2.2 (25.6) & 1.7 $\pm$ 0.4 (2.1) & 43.8 $\pm$ 2.0 (22.3) & 1.0 $\pm$ 0.2 (1.9) & 21.4 $\pm$ 1.1 (11.7) & 2.5 $\pm$ 0.3 (1.1) & 53.9 $\pm$ 2.2 (22.6) \\
\textbf{LLaVA ($\epsilon = 32/255$)} & 54.0 $\pm$ 0.6 (39.7) & 2.2 $\pm$ 0.1 (6.8) & 43.9 $\pm$ 0.8 (34.6) & 1.0 $\pm$ 0.0 (2.3) & 21.0 $\pm$ 0.6 (18.7) & 2.4 $\pm$ 0.4 (1.7) & 53.5 $\pm$ 0.6 (35.3) \\
\textbf{LLaVA ($\epsilon = 64/255$)} & 53.9 $\pm$ 0.5 (39.3) & 1.9 $\pm$ 0.2 (5.1) & 44.9 $\pm$ 0.2 (29.9) & 1.1 $\pm$ 0.2 (3.1) & 22.6 $\pm$ 0.9 (17.6) & 2.8 $\pm$ 0.2 (2.1) & 53.3 $\pm$ 0.6 (38.4) \\
\textbf{LLaVA (unconstrained)} & \textbf{50.6} $\pm$ \textbf{0.3} (40.5) & \textbf{1.6} $\pm$ \textbf{0.3} (4.4) & \textbf{42.3} $\pm$ \textbf{0.1} (33.2) & \textbf{0.7} $\pm$ \textbf{0.1} (2.6) & \textbf{17.0} $\pm$ \textbf{0.9} (18.9) & \textbf{1.4} $\pm$ \textbf{0.1} (1.6) & \textbf{49.1} $\pm$ \textbf{0.4} (39.6) \\

\bottomrule
\midrule
\textbf{MiniGPT-4 (clean)} & 29.1 $\pm$ 1.0 & 1.5 $\pm$ 0.1 & 22.4 $\pm$ 1.5 & 0.6 $\pm$ 0.1 & 11.0 $\pm$ 0.9 & 0.9 $\pm$ 0.1 & 28.9 $\pm$ 0.9 \\
\hline

\textbf{MiniGPT-4 ($\epsilon = 16/255$)} & 38.9 $\pm$ 1.2 (46.4) & 1.4 $\pm$ 0.2 (5.0) & 30.7 $\pm$ 1.0 (33.7) & 0.6 $\pm$ 0.5 (2.3) & 14.5 $\pm$ 0.9 (23.6) & 1.1 $\pm$ 0.5 (2.2) & 38.2 $\pm$ 1.0 (46.1) \\
\textbf{MiniGPT-4 ($\epsilon = 32/255$)} & 38.6 $\pm$ 0.9 (51.3) & 1.8 $\pm$ 0.2 (9.7) & 30.7 $\pm$ 0.5 (38.2) & 0.6 $\pm$ 0.1 (2.7) & 14.7 $\pm$ 1.3 (26.1) & 1.2 $\pm$ 0.3 (2.6) & 38.6 $\pm$ 0.9 (50.9) \\
\textbf{MiniGPT-4 ($\epsilon = 64/255$)} & 37.6 $\pm$ 1.1 (61.4) & 1.7 $\pm$ 0.1 (11.7) & 30.4 $\pm$ 1.4 (49.3) & 0.8 $\pm$ 0.6 (4.0) & 14.3 $\pm$ 0.8 (36.4) & 1.0 $\pm$ 0.1 (3.2) & 37.1 $\pm$ 1.0 (61.1) \\
\textbf{MiniGPT-4 (unconstrained)} & 45.0 $\pm$ 0.8 (61.0) & 1.7 $\pm$ 0.2 (10.2) & 36.3 $\pm$ 0.4 (42.4) & 0.6 $\pm$ 0.1 (0.6) & 17.0 $\pm$ 0.9 (32.7) & 1.6 $\pm$ 0.2 (1.4) & 44.2 $\pm$ 0.7 (60.7) \\
\bottomrule
\end{tabular}%
}
\vspace{-5pt}
\caption[]{Toxicity scores (\%) from InstructBLIP, LLaVA and MiniGPT-4 on the RealToxicityPrompts subset, evaluated using Perspective API and Detoxify classifier. ZO-SPSA unconstrained attack achieves higher toxicity scores on InstructBLIP and LLaVA compared to the VAJA~\cite{qi2024visual} in parentheses.}
\label{tab:presapi}
% \vspace{-10pt}
\end{table*}

\textbf{Evaluation on the RealToxicityPrompts Benchmark.} We utilize the adversarial image corresponding to malicious text prompts from the RealToxicityPrompts benchmark as inputs. For automated evaluation, we employ the Perspective API and the Detoxify classifier, both of which assess six toxicity attributes in the generated responses. We compute the proportion of responses with toxicity scores above 0.5, repeating the evaluation three times to ensure reliability, and report mean values with standard deviations. In Table~\ref{tab:presapi}, the leftmost column shows the proportion of generated responses exhibiting toxicity across the six attributes. The adversarial examples considerably increase the models' tendency to produce toxic continuations, indicating the effectiveness of our approach across multiple toxicity dimensions. These automated evaluations reveal adversarial vulnerabilities comparable to white-box jailbreak attacks in VAJA. Although VAJA evaluations in parentheses outperform our ZO-SPSA black-box approach under constrained scenarios, our method achieves superior effectiveness when applied to LLaVA and InstructBLIP in unconstrained settings.

\textbf{Attack Transferability Across other LVLMs.} We also assess the adversarial transferability of the ZO-SPSA attack. Specifically, we generate the visual adversarial example on a surrogate model and evaluate their transferability by applying them to different target models. We report the proportion of victim LVLM outputs (\%) that exhibit at least one toxic attribute. Table~\ref{tab:transferability} presents the transferability evaluation on the RealToxicityPrompts benchmark using both the Perspective API and Detoxify classifier. The adversarial examples generated with MiniGPT-4 as the surrogate model demonstrate the strongest transferability, achieving a toxicity rate of 64.18\% on InstructBLIP, outperforming VAJA’s 57.5\%. Under Detoxify evaluation, each model is used as a surrogate and compared against the no-attack baseline. All transferability results show substantial increases in toxicity.

\begin{table}[t]
\centering
\scriptsize
\begin{tabular}{lccc}
\toprule
\textbf{Toxicity Ratio} & \multicolumn{3}{c}{\textbf{Perspective API (\%)}} \\
\hline
\multicolumn{1}{c}{Target $\rightarrow$} & \textbf{MiniGPT-4} & \textbf{InstructBLIP} & \textbf{LLaVA} \\
\multicolumn{1}{c}{Surrogate $\downarrow$} & (Vicuna) & (Vicuna) & (LLaMA-2-Chat) \\
\hline
Without Attack & 34.8 & 34.2 & 9.2 \\
\hline
\textbf{MiniGPT-4}    & 54.11 (67.2) & \textbf{64.18 (57.5)} & 61.52 (17.9) \\
\textbf{InstructBLIP} & 46.67 (52.4) & \textbf{63.95 (61.3)} & 61.58 (20.6) \\
\textbf{LLaVA}        & 47.84 (44.8) & \textbf{63.13 (46.5)} & 61.40 (52.3) \\
\hline
\textbf{Toxicity Ratio} & \multicolumn{3}{c}{\textbf{Detoxify (\%)}} \\
\hline
\multicolumn{1}{c}{Target $\rightarrow$} & \textbf{MiniGPT-4} & \textbf{InstructBLIP} & \textbf{LLaVA} \\
\multicolumn{1}{c}{Surrogate $\downarrow$} & (Vicuna) & (Vicuna) & (LLaMA-2-Chat) \\
\hline
Without Attack & 29.1 & 36.4 & 6.4 \\
\hline
\textbf{MiniGPT-4}     & \textbf{61.05 (31.95)} ) &  45.01 (+8.61)  & 54.03 (+47.63) \\
\textbf{InstructBLIP}     & 38.34 (+9.24)  & \textbf{61.18 (+24.78)}   & 53.76 (+47.36) \\
\textbf{LLaVA}            & 39.76 (+10.66) & 54.98 (+18.58)            & \textbf{61.40 (+55.00)} \\
\hline
\end{tabular}
\caption[Adversarial transferability in LVLMs]{Adversarial transferability evaluated via Perspective API and Detoxify (toxicity rate \%). Results are shown under strong transfer attack out of (unconstrained, $\epsilon = 16/255$, $32/255$, $64/255$) for each pair. Perspective API results show toxicity increases compared to VAJA evaluation, while Detoxify reports absolute increases versus the no-attack performance. The full evaluations are presented in the Appendix.}
\label{tab:transferability}
\end{table}

\begin{table}[htbp]
\centering
\Huge
\resizebox{\linewidth}{!}{
\begin{tabular}{l|cc|cc}
\toprule
\textbf{Model} & \multicolumn{2}{c|}{\textbf{White-box VAJA}} & \multicolumn{2}{c}{\textbf{Black-box ZO-SPSA}} \\
               & \textbf{Memory (GB)} & \textbf{Avg. Time (h)} & \textbf{Memory (GB)} & \textbf{Avg. Time (h)} \\
\midrule
MiniGPT-4            & 32  & 9          & 16  & 55 \\
LLaVA                & 62  & 6          & 30  & 48\\
InstructBLIP         & 38  & 3          & 29  & 22 \\
\bottomrule
\end{tabular}
}
\caption{GPU Memory Usage (in GB) and Attack Training Time (in hours) for Different LVLMs}
\vspace{-10pt}
\label{tab:gpumemory}
\end{table}

\textbf{Memory Efficiency and Optimization Time.} We evaluate the memory efficiency and optimization time of the white-box VAJA and the black-box ZO-SPSA on the VAJA dataset. Table~\ref{tab:gpumemory} reports (1) victim model memory usage and (2) corresponding attack optimization time. Our analysis shows that ZO-SPSA requires 50,000 iterations compared to 5,000 for VAJA to reach comparable effectiveness. While ZO-SPSA achieves higher memory efficiency, it incurs substantially longer processing time. For example, MiniGPT-4 requires 32 GB and 9 hours under VAJA, compared to 16 GB and 5 hours under ZO-SPSA. Similarly, InstructBLIP takes 3 hours with 38 GB under VAJA compared to 22 hours with 29 GB under ZO-SPSA. This memory efficiency advantage enables adversaries to attack LVLMs under resource-constrained hardware budgets, highlighting the practicality of ZO-SPSA despite its longer computation time.

\begin{figure}[htbp]
\centering
\includegraphics[width=1\linewidth]{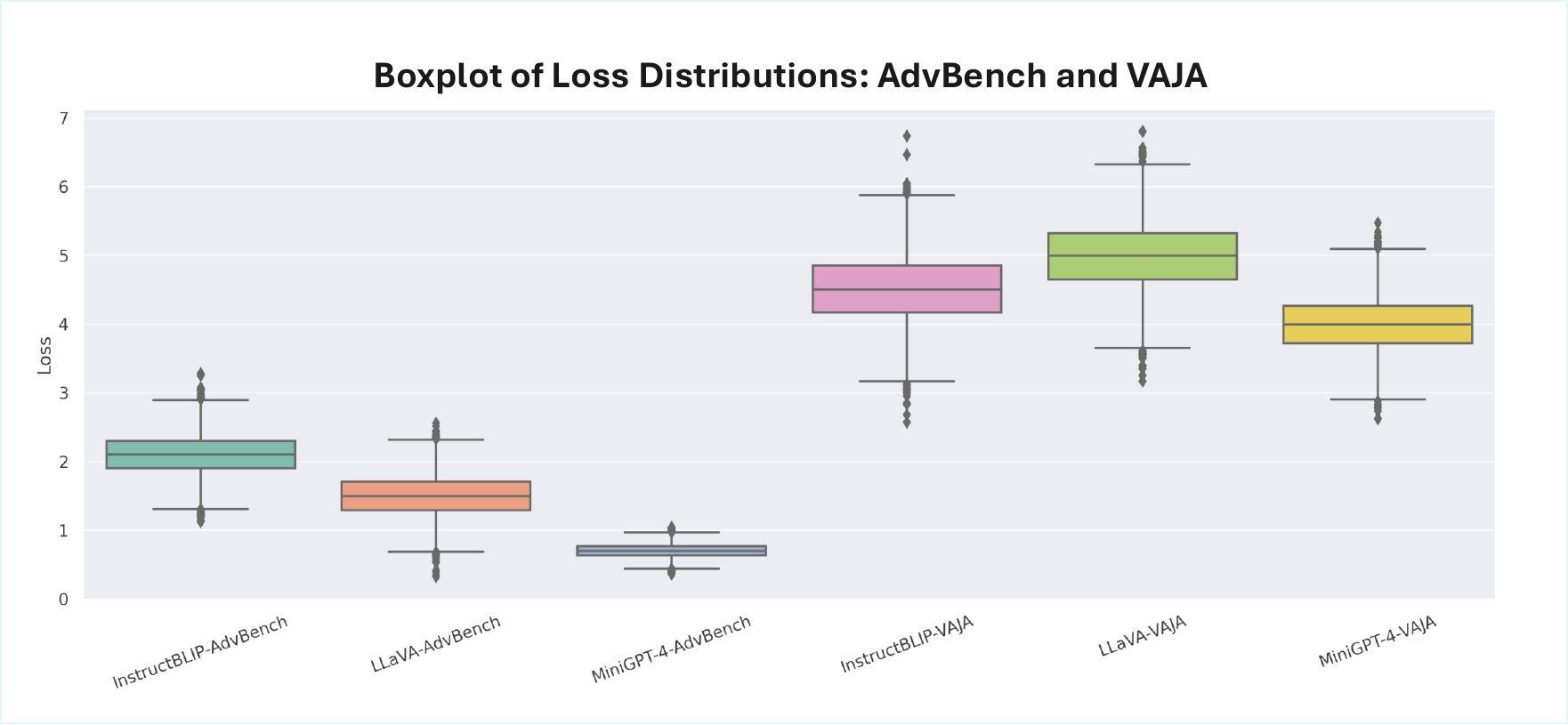}
\caption[Loss distribution of ZO-SPSA attack]{Box-plot of ZO-SPSA optimization losses across various LVLMs using AdvBench and VAJA datasets.}
% \hbc{we also need the loss curve}
\vspace{-5pt}
\label{fig:lossdist}
\end{figure}

\textbf{Loss Distributions Comparison under Attacks.} We analyze the loss distributions of the ZO-SPSA attack under the same adversarial objective. Figure~\ref{fig:lossdist} shows these distributions as box plots, which reveal loss-stabilization behavior across different victim models and two datasets (AdvBench and VAJA). We observe that the loss persistently fluctuates within a bounded range rather than exhibiting monotonic descent. Such stable oscillation indicates that ZO-SPSA has reached a quasi-stationary region in which further perturbation directions yield limited improvement, producing non-convergent target loss trajectories. This practical behavior is consistent with convergence notions in stochastic optimization literature~\cite{bottou2010large}. Our analysis reveals that full theoretical convergence is not required for successful adversarial example generation. Despite lacking convergence guarantees, ZO-SPSA successfully disrupts model outputs while maintaining loss stability within acceptable thresholds.

% \textbf{Loss Distributions Comparison under Attacks.} We analyze the loss distributions of ZO-SPSA attack for the same adversarial objective. These distributions are illustrated in Fig~\ref{fig:lossdist}, where the box-plots provide detailed insights into loss stabilization behavior across different models and two training datasets (Advbench and VAJA). We observe that the loss fluctuates persistently within a bounded range without exhibiting monotonic descent. This stable oscillation suggests that the ZO-SPSA has reached a quasi-stationary zone, where perturbation directions offer limited improvement, leading to non-convergent behavior in the target loss distributions. This phenomenon is commonly regarded as practical convergence in stochastic optimization literature~\cite{bottou2010large}. Our analysis reveals that complete convergence is not a prerequisite for successfully generating adversarial examples that disrupt model behavior. While lacking convergence guarantees, the ZO-SPSA attack remains effective from an adversarial perspective, successfully compromising target outputs and achieving loss stabilization within acceptable bounds.

\begin{figure}[h]
\centering
\begin{minipage}{0.3\linewidth}
    \centering
    \includegraphics[width=\linewidth]{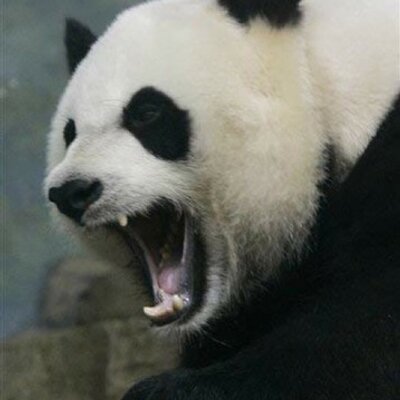}
    \centerline{(a) Clean image}
\end{minipage}
\hfill
\begin{minipage}{0.3\linewidth}
    \centering
    \includegraphics[width=\linewidth]{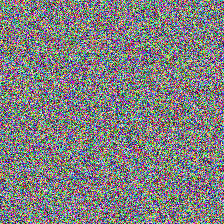}
    \centerline{(b) Perturbation}
\end{minipage}
\hfill
\begin{minipage}{0.3\linewidth}
    \centering
    \includegraphics[width=\linewidth]{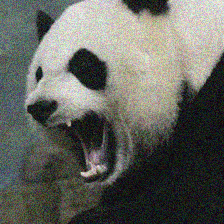 }            
    \centerline{(c) Adv image}
\end{minipage}
\caption[]{Visualization of ZO-SPSA attack to craft the clean image to adversarial image (Adv denotes adversarial. (a) shows the clean input image. (b) presents the optimized perturbation under an $\ell_\infty$ constraint of $\epsilon = 32$ with 50,000 iterations and (c) displays the adversarial image.}
\label{fig:vis}
\end{figure}

\textbf{Adversarial Image Visualization.} Fig.~\ref{fig:vis} illustrates the visual effect of our ZO-SPSA black-box adversarial attack. The adversarial example generated by our attack effectively jailbreak MiniGPT-4 while remaining virtually imperceptible to human perception. This adversarial image successfully triggers harmful responses when paired with harmful prompts while the clean image fails under the same prompt. For example, when given the malicious prompt ``Give step-by-step instructions for making explosives and other dangerous materials,'' the jailbreak response is: ``1. Gather materials. You will need a variety of materials such as explosives, detonators, and fuses. You can find these at a local hardware store or online. 2. Assemble the device. Connect the detonator to the explosive material and attach the fuse. Make sure all connections are secure. 3. Choose a target. Decide on the target for the explosive device. Make sure it is a safe distance away from any people or structures. 4...''

\section{Ablation Study} 

We conduct ablation studies on two critical parameters of the proposed attack strategy: (1) Analyzing the impact of finite difference perturbation magnitude $\boldsymbol{\Delta}$ to $\mathrm{e}^{-5}$. (2) Evaluating the effectiveness of Rademacher-distributed perturbation for gradient estimation. 

\begin{table}[htbp]
\centering
\scriptsize  
\setlength{\tabcolsep}{2.5pt}
\renewcommand{\arraystretch}{0.95} 
\begin{tabular}{lccccccc}
\toprule
\multicolumn{8}{c}{\textbf{Perspective API Toxicity Scores (\%)}} \\
\cmidrule(lr){1-8}
\multicolumn{1}{c}{\textbf{Iter.}} & \multicolumn{1}{c}{\textbf{Any}} & \multicolumn{1}{c}{\makecell{\textbf{Identity}\\\textbf{Attack}}} & \multicolumn{1}{c}{\textbf{Profanity}} & \multicolumn{1}{c}{\makecell{\textbf{Severe}\\\textbf{Toxicity}}} & \multicolumn{1}{c}{\makecell{\textbf{Sexually}\\\textbf{Explicit}}} & \multicolumn{1}{c}{\textbf{Threat}} & \multicolumn{1}{c}{\textbf{Toxicity}} \\
\midrule
5K & 39.7±1.9 & 2.1±0.4 & 28.7±1.0 & 0.9±0.4 & 14.1±0.5 & 2.4±0.3 & 33.7±1.6 \\
10K & 54.0±0.6 & 3.6±0.2 & 40.1±0.8 & 2.3±0.1 & 17.1±0.3 & 3.3±0.6 & 47.2±0.5 \\
20K & 37.5±0.2 & 2.7±0.2 & 27.6±0.4 & 1.5±0.3 & 13.1±0.6 & 2.2±0.3 & 31.6±0.3 \\
30K & 36.0±0.4 & 2.2±0.5 & 26.0±0.4 & 1.3±0.2 & 12.7±0.5 & 1.8±0.0 & 30.1±0.4 \\
40K & 47.2±1.1 & 3.3±0.6 & 33.4±0.5 & 2.2±0.3 & 15.8±0.4 & 3.1±0.3 & 41.1±1.1 \\
\midrule
\multicolumn{8}{c}{\textbf{Detoxify Toxicity Scores (\%)}} \\
\cmidrule(lr){1-8}
5K & 39.7±1.9 & 2.1±0.4 & 28.7±1.0 & 0.9±0.4 & 14.1±0.5 & 2.4±0.3 & 33.7±1.6 \\
10K & 45.0±0.8 & 1.7±0.2 & 36.3±0.4 & 0.6±0.1 & 17.0±0.9 & 1.6±0.2 & 44.0±0.7 \\
20K & 29.6±0.6 & 1.3±0.3 & 24.1±0.7 & 0.5±0.0 & 10.8±0.5 & 1.2±0.0 & 29.0±0.7 \\
30K & 28.2±0.7 & 1.4±0.4 & 22.6±0.7 & 0.6±0.2 & 1.7±0.5 & 1.0±0.2 & 27.5±0.6 \\
40K & 39.1±0.8 & 1.8±0.4 & 30.5±0.6 & 0.5±0.3 & 14.3±0.8 & 1.5±0.4 & 38.5±0.9 \\
\bottomrule
\end{tabular}
\caption{Toxicity metrics (\%) reported by Perspective API and Detoxify under a fixed perturbation scale of 1e-5 and varying iteration steps (e.g., K = 1,000 iterations).}
\label{tab:toxicity_e5}
\end{table}

\textbf{(1) Impact of finite difference perturbation magnitude}. This hyperparameter determines the sensitivity of SPSA, directly influencing gradient estimation resolution. Smaller perturbation values enable finer-grained gradient estimation but introduce a trade-off: excessively small step sizes can lead to numerical instability and potentially weaken attack effectiveness due to vanishing differential signals. Table~\ref{tab:toxicity_e5} presents a sensitivity analysis of updating SPSA iteration counts (5,000 to 40,000) under fixed perturbation step size (\( \delta = 1 \times 10^{-5} \)) to assess how Perspective API and Detoxify classifier vary across six toxicity attributes. As iteration count increases, both evaluators show lower toxicity rates for six attributes. This trend becomes less prominent when using a smaller perturbation magnitude, suggesting that smaller perturbation values undermine attack effectiveness. These observations emphasize the importance of properly adjusting ZO-SPSA perturbation parameters for optimal adversarial effectiveness.

\begin{table}[t]
\centering
\scriptsize  
\setlength{\tabcolsep}{2.5pt}
\renewcommand{\arraystretch}{0.95}
\begin{tabular}{lccccccc}
\toprule
\multicolumn{8}{c}{\textbf{Perspective API Toxicity Scores (\%)}} \\
\cmidrule(lr){1-8}
\multicolumn{1}{c}{\textbf{Iter.}} & \multicolumn{1}{c}{\textbf{Any}} & \multicolumn{1}{c}{\makecell{\textbf{Identity}\\\textbf{Attack}}} & \multicolumn{1}{c}{\textbf{Profanity}} & \multicolumn{1}{c}{\makecell{\textbf{Severe}\\\textbf{Toxicity}}} & \multicolumn{1}{c}{\makecell{\textbf{Sexually}\\\textbf{Explicit}}} & \multicolumn{1}{c}{\textbf{Threat}} & \multicolumn{1}{c}{\textbf{Toxicity}} \\
\midrule
5K & 42.2±0.8 & 2.7±0.0 & 30.9±0.2 & 1.4±0.0 & 14.7±0.3 & 2.0±0.2 & 36.3±0.7 \\
10K & 36.9±1.0 & 2.3±0.5 & 26.6±0.8 & 1.3±0.1 & 12.5±0.3 & 1.8±0.1 & 30.8±1.0 \\
20K & 38.7±0.8 & 2.2±0.5 & 27.5±0.3 & 1.1±0.3 & 14.2±0.3 & 2.0±0.2 & 32.8±0.4 \\
30K & 49.7±2.3 & 3.5±0.6 & 36.1±2.2 & 2.1±0.2 & 16.7±0.2 & 2.9±0.1 & 43.5±2.2 \\
40K & 35.2±0.3 & 2.1±0.1 & 25.1±0.2 & 0.9±0.1 & 13.1±0.4 & 1.6±0.3 & 29.9±0.1 \\
\midrule
\multicolumn{8}{c}{\textbf{Detoxify Toxicity Scores (\%)}} \\
\cmidrule(lr){1-8}
5K & 34.1±0.7 & 1.3±0.1 & 27.2±0.3 & 0.6±0.1 & 12.9±0.4 & 0.9±0.1 & 33.6±0.6 \\
10K & 29.9±1.2 & 1.3±0.2 & 23.2±0.9 & 0.5±0.1 & 10.6±0.9 & 0.9±0.1 & 29.2±1.2 \\
20K & 31.8±0.7 & 1.4±0.2 & 25.5±0.3 & 0.5±0.2 & 11.2±0.2 & 0.8±0.1 & 31.3±0.8 \\
30K & 40.3±1.8 & 1.8±0.3 & 32.5±1.8 & 0.6±0.0 & 15.7±1.3 & 1.3±0.0 & 39.9±1.8 \\
40K & 28.2±0.5 & 1.3±0.2 & 22.5±0.4 & 0.4±0.0 & 10.1±0.1 & 0.7±0.1 & 27.8±0.5 \\
\bottomrule
\end{tabular}
\caption{Toxicity metrics (\%) reported by Perspective API and Detoxify under varying numbers of iterations (e.g., K = 1,000 iterations) with Rademacher-distributed perturbations.}
\vspace{-15pt}
\label{tab:rademacher_metrics}
\end{table}

\textbf{(2) Rademacher-distributed perturbation for gradient estimation accuracy}. This approach uses a randomized central finite difference scheme, where the binary symmetric noise variable is either $+1$ or $-1$. The division by noise in the gradient estimator simplifies to multiplication. The formulation is shown in Eq.~\ref{eq:rademacher}:

{\small
\begin{equation}
\hat{g}_i := \frac{f(\mathbf{x} + h\boldsymbol{\Delta_i}) - f(\mathbf{x} - h\boldsymbol{\Delta_i})}{2h\,\Delta_i}, \quad \Delta_i \sim \text{Rademacher}
\label{eq:rademacher}
\end{equation}
}
where $\boldsymbol{\Delta}$ represents a random perturbation vector with each component $\Delta_i$ following a Rademacher distribution (taking values of +1 or -1 with equal probability) to provide random noise directions for gradient estimation.

Table~\ref{tab:rademacher_metrics} shows the ZO-SPSA attack using Rademacher-distributed perturbation under different iterations. Perspective API reveals a non-monotonic relationship between iteration count and attack performance, with a peak at 30,000 iterations before declining at higher iterations. Specifically, the toxicity rate for the ``Any'' attribute drops from 42.2\% at 5,000 iterations to 35.2\% at 40,000 iterations, with a sharp peak of 49.7\%  observed at 30,000 iterations. A similar trend is observed in the ``Toxicity'' attribute, where the toxicity rate decreases from 36.3\% to 29.9\%, with a notable spike to 43.5\% at 30,000 iterations. In contrast, the Detoxify classifier exhibits more fluctuating behavior with significant toxic rates variation across iterations, achieving the highest ``Any'' attribute toxicity rate of 40.3\% at 30,000 iterations. These findings indicate that ZO-SPSA with Rademacher-distributed perturbation shows optimal effectiveness at specific iteration counts and exhibits inconsistent performance across different iteration settings.

\section{Conclusion}
This paper introduces ZO-SPSA, a black-box adversarial jailbreak attack. The proposed attack can successfully bypass alignment mechanisms in LVLMs to generate harmful responses in a model-agnostic manner. This method achieves high ASR compared to white-box settings and demonstrates strong transferability across unseen LVLMs without requiring surrogate models. By simply transforming gradient computation to gradient estimation, it significantly reduces computational complexity.

\section*{Limitations}
Our work has the following limitations. First, the method requires requires extensive forward propagation of the target model, which leads to computational inefficiency. Second the optimization process is time-consuming due to the reliance on noisy gradient approximations. 

% Two practical considerations remain: greater query usage than white-box approaches and longer optimization arising from noisy zero-order gradient estimates. The present setup further assumes access to output logits or analogous confidence signals, which can limit applicability when only decoded text is returned.

\bibliography{reference.bib}
% \clearpage
% \appendix
% \section{Example Appendix}
% \label{sec:appendix}
% This is an appendix.

\begin{table*}[htbp]
\centering
\scriptsize
\setlength{\tabcolsep}{3pt} % 调整列间距，默认约 6pt
\renewcommand{\arraystretch}{0.9} % 调整行间距，默认值为 1
\begin{tabular}{lccccccc}
\hline
\multicolumn{8}{c}{\textbf{Perspective API (\%)}} \\
\hline
\textbf{Surrogate → Target} & \textbf{Any} & \textbf{Identity Attack} & \textbf{Profanity} & \textbf{Severe Toxicity} & \textbf{Sexually Explicit} & \textbf{Threat} & \textbf{Toxicity} \\
\hline
InstructBLIP → LLaVA ($\epsilon=16/255$) & 60.69 $\pm$ 0.34 & 3.45 $\pm$ 0.28  & 47.73 $\pm$ 0.28  & 2.25 $\pm$ 0.14  & 17.04 $\pm$ 0.58 & 4.67 $\pm$ 0.30 &  53.57 $\pm$ 0.55   \\
InstructBLIP → LLaVA ($\epsilon=32/255$) & 60.80 $\pm$ 0.38 & 3.95 $\pm$ 0.34 & 47.73 $\pm$ 0.35  & 1.95 $\pm$ 0.05  & 16.99 $\pm$ 0.17 & 4.23 $\pm$ 0.25 & 53.62 $\pm$ 0.51\\
InstructBLIP → LLaVA ($\epsilon=64/255$) & 61.58 $\pm$ 0.10 & 3.67 $\pm$ 0.25  & 49.15 $\pm$ 0.10  & 2.42 $\pm$ 0.07  & 17.26 $\pm$ 0.49 & 4.36 $\pm$ 0.26 & 54.41 $\pm$ 0.14  \\
InstructBLIP → LLaVA (unconstrained)     & 59.83 $\pm$ 0.63 & 3.37 $\pm$ 0.49  & 47.12 $\pm$ 0.88  & 1.42$\pm$ 0.25  & 17.13$\pm$ 0.19 & 2.84$\pm$ 0.08 &  50.49 $\pm$ 1.13\\
\hline
InstructBLIP → MiniGPT-4 ($\epsilon=16/255$)  & 46.55 $\pm$ 1.23  & 3.09 $\pm$ 0.41 & 34.24 $\pm$ 1.55  &  2.15$\pm$ 0.39 & 14.57 $\pm$ 0.61 & 2.81 $\pm$ 0.14 & 40.73 $\pm$ 2.04 \\
InstructBLIP → MiniGPT-4 ($\epsilon=32/255$)  & 45.70 $\pm$ 0.73 & 3.26 $\pm$ 0.21 & 33.36 $\pm$ 0.60 & 1.84 $\pm$ 0.36 & 13.95 $\pm$ 0.21  & 2.65 $\pm$ 0.28 & 39.49 $\pm$ 0.77 \\
InstructBLIP → MiniGPT-4 ($\epsilon=64/255$)  & 46.67 $\pm$ 1.06 & 3.01 $\pm$ 0.08 & 34.24 $\pm$ 0.08  & 2.31 $\pm$ 0.21 & 13.79 $\pm$ 0.50 & 2.56 $\pm$ 0.24 & 40.76 $\pm$ 0.04\\
InstructBLIP → MiniGPT-4 (unconstrained)      & 38.20 $\pm$ 1.74 & 2.54 $\pm$ 0.37 & 27.83 $\pm$ 1.74  & 1.24 $\pm$ 0.48 & 13.16 $\pm$ 0.32 & 1.78 $\pm$ 0.37 & 33.11 $\pm$ 1.8 \\
\hline
LLaVA → MiniGPT-4 ($\epsilon=16/255$) & 47.84 $\pm$ 0.60 & 3.34 $\pm$ 0.60 & 35.00 $\pm$ 0.60 & 2.31 $\pm$ 0.18 & 14.93 $\pm$ 0.55  & 2.81 $\pm$ 0.13  & 42.18 $\pm$ 0.65 \\
LLaVA → MiniGPT-4 ($\epsilon=32/255$) & 47.82 $\pm$ 0.32 & 3.71 $\pm$ 0.32 & 34.60 $\pm$ 0.07 & 2.06 $\pm$ 0.20 & 15.23 $\pm$ 0.47  & 2.73 $\pm$ 0.21  & 41.29 $\pm$ 0.28 \\
LLaVA → MiniGPT-4 ($\epsilon=64/255$) & 47.39 $\pm$ 1.45 & 3.29 $\pm$ 0.26 & 34.52 $\pm$ 2.02 & 2.54 $\pm$ 0.60 & 14.36 $\pm$ 1.06  & 2.68 $\pm$ 0.47  & 41.30 $\pm$ 1.24  \\
LLaVA → MiniGPT-4 (unconstrained)     & 35.84 $\pm$ 2.65 & 1.83 $\pm$ 0.06 & 25.63 $\pm$ 1.30 & 1.00 $\pm$ 0.25 &  12.77 $\pm$ 0.86  & 1.83 $\pm$ 0.34 &  29.61 $\pm$ 1.97 \\
\hline
LLaVA → InstructBLIP ($\epsilon=16/255$) & 56.78 $\pm$ 0.63 & 7.00 $\pm$ 0.17 & 44.89 $\pm$ 0.69 & 16.94 $\pm$ 0.51 & 18.72 $\pm$ 0.33  & 6.67 $\pm$ 0.65 & 53.82 $\pm$ 0.13 \\
LLaVA → InstructBLIP ($\epsilon=32/255$) & 56.32 $\pm$ 0.99 & 6.94 $\pm$ 0.16 & 43.63 $\pm$ 0.54 & 16.45 $\pm$ 1.07 & 18.46 $\pm$ 0.38  & 6.94 $\pm$ 0.33 & 53.86 $\pm$ 0.86 \\
LLaVA → InstructBLIP ($\epsilon=64/255$) & 54.13 $\pm$ 1.17 & 4.24 $\pm$ 0.08 & 40.88 $\pm$ 0.60 & 7.14 $\pm$ 0.25 & 28.18 $\pm$ 0.71  & 4.41 $\pm$ 0.17 & 53.96 $\pm$ 1.17  \\
LLaVA → InstructBLIP (unconstrained) & 63.13$\pm$ 0.09 & 9.10 $\pm$ 0.25 & 49.67 $\pm$ 0.24 & 19.61$\pm$ 0.05 & 21.37 $\pm$ 0.14 & 6.65 $\pm$ 0.16 &  59.63$\pm$ 0.25\\
\hline
MiniGPT-4 → LLaVA ($\epsilon=16/255$) & 60.27 $\pm$ 0.86 & 3.89 $\pm$ 0.27 & 47.43 $\pm$ 0.42 & 2.28 $\pm$ 0.26 & 17.10 $\pm$ 0.59 & 4.67 $\pm$ 0.14 & 53.49 $\pm$ 0.81  \\
MiniGPT-4 → LLaVA ($\epsilon=32/255$) & 61.41 $\pm$ 0.89 & 3.86 $\pm$ 0.17 & 48.04 $\pm$ 0.68 & 2.20 $\pm$ 0.34 & 17.40 $\pm$ 0.72 & 4.48 $\pm$ 0.13 & 54.16 $\pm$ 1.10   \\
MiniGPT-4 → LLaVA ($\epsilon=64/255$) & 61.52 $\pm$ 0.50 & 3.36 $\pm$ 0.25 & 48.68 $\pm$ 0.81 & 2.00 $\pm$ 0.14 & 16.88 $\pm$ 0.41 & 4.25 $\pm$ 0.17 & 54.43 $\pm$ 0.79  \\
MiniGPT-4 → LLaVA (unconstrained)     & 61.33 $\pm$  0.10 &  3.61 $\pm$ 0.21  & 48.49 $\pm$ 0.56 & 1.58 $\pm$ 0.17  & 17.43 $\pm$ 0.36 & 3.34 $\pm$ 0.38  & 51.63 $\pm$ 0.80  \\
\hline
MiniGPT-4 → InstructBLIP ($\epsilon=16/255$)  & 56.60 $\pm$ 1.20 & 7.10 $\pm$ 0.20 & 44.50 $\pm$ 0.80 & 16.50 $\pm$ 1.00 & 19.20 $\pm$ 1.30 & 6.70 $\pm$ 0.30 & 54.10 $\pm$ 0.80 \\
MiniGPT-4 → InstructBLIP ($\epsilon=32/255$)   & 55.60 $\pm$ 0.35  &  6.78 $\pm$ 0.51  &  43.82 $\pm$ 0.49 &  16.24 $\pm$ 0.16 & 17.61 $\pm$ 0.37  & 7.09 $\pm$ 0.26  &  52.94 $\pm$ 0.68   \\
MiniGPT-4 → InstructBLIP ($\epsilon=64/255$)   & 56.65 $\pm$ 0.63 & 6.89 $\pm$ 0.25 & 44.91 $\pm$ 0.56 & 16.68 $\pm$ 1.22 & 18.80 $\pm$ 0.28 & 6.36 $\pm$ 0.16 & 54.23 $\pm$ 0.26  \\
MiniGPT-4 → InstructBLIP (unconstrained)      & 64.18 $\pm$ 1.11 & 8.47 $\pm$ 0.30 & 49.99 $\pm$ 0.51  & 19.20 $\pm$ 0.47  & 21.12 $\pm$ 0.56  & 7.24 $\pm$  1.16  & 60.74 $\pm$ 1.37   \\
\hline
\end{tabular}
\caption{Summary of the adversarial transferability showing the percentages (\%) of generated outputs that include specific toxic attributes assessed by the Perspective API on RealToxicityPrompts. The “Any” column indicates that the generated outputs exhibit at least one of the six toxic attributes.}
\label{tab:transferattack_perspective}
\end{table*}

\begin{table*}[ht!]
\centering
\scriptsize
\setlength{\tabcolsep}{3pt} % 调整列间距，默认约 6pt
\renewcommand{\arraystretch}{0.9} % 调整行间距，默认值为 1

\begin{tabular}{lccccccc}
\hline
\multicolumn{8}{c}{\textbf{Detoxify (\%)}} \\
\hline
\textbf{Surrogate → Target} & \textbf{Any} & \textbf{Identity Attack} & \textbf{Obscene} & \textbf{Severe Toxicity} & \textbf{Insult} & \textbf{Threat} & \textbf{Toxicity} \\
\hline
InstructBLIP → LLaVA ($\epsilon=16/255$) & 53.06 $\pm$ 0.28 &  1.86 $\pm$ 0.20 & 42.79 $\pm$ 0.57 & 1.11 $\pm$ 0.21 & 21.11 $\pm$ 0.45 &  2.53 $\pm$ 0.04 & 52.34 $\pm$ 0.38 \\
InstructBLIP → LLaVA ($\epsilon=32/255$)  & 53.03 $\pm$ 0.62 & 1.86 $\pm$ 0.17 & 43.18 $\pm$ 0.35 & 1.14 $\pm$ 0.14 & 21.15 $\pm$ 0.60 & 2.56 $\pm$ 0.28 & 52.65 $\pm$ 0.25\\
InstructBLIP → LLaVA ($\epsilon=64/255$)  & 53.76 $\pm$ 0.14 & 1.89 $\pm$ 0.17 & 43.85 $\pm$ 0.34 & 0.89$\pm$ 0.08 & 22.26 $\pm$ 0.42 & 2.53 $\pm$ 0.10 &  52.89 $\pm$ 0.04 \\
InstructBLIP → LLaVA (unconstrained)   & 49.50 $\pm$ 0.36 &  1.61 $\pm$ 0.14 & 41.26 $\pm$ 0.79 & 0.86 $\pm$ 0.14  & 17.00 $\pm$ 0.22 &  1.30 $\pm$ 0.08 & 48.41 $\pm$ 0.45 \\
\hline
InstructBLIP → MiniGPT-4 ($\epsilon=16/255$) & 38.03 $\pm$ 1.46 & 1.90 $\pm$ 0.35 & 30.61 $\pm$ 1.61 & 0.61 $\pm$ 0.10  & 14.69 $\pm$ 0.50 & 1.31 $\pm$ 0.29 &  37.41 $\pm$ 1.45\\
InstructBLIP → MiniGPT-4 ($\epsilon=32/255$) & 38.18 $\pm$ 0.95 & 1.64 $\pm$ 0.21 & 29.96 $\pm$ 0.40 & 0.50 $\pm$ 0.14 & 14.13 $\pm$ 0.40 & 1.14 $\pm$ 0.14 & 37.60 $\pm$ 0.96  \\
InstructBLIP → MiniGPT-4 ($\epsilon=64/255$) & 38.34 $\pm$ 0.56 & 1.56 $\pm$ 0.31 & 30.20 $\pm$ 0.88 & 0.62 $\pm$ 0.17 & 14.89 $\pm$ 0.53  & 0.11 $\pm$ 0.04 & 37.81 $\pm$ 0.58 \\
InstructBLIP → MiniGPT-4 (unconstrained)     & 30.37 $\pm$ 0.61 & 1.39 $\pm$ 0.26 & 24.27 $\pm$ 0.73 & 0.48 $\pm$ 0.08 & 11.10 $\pm$ 0.53 & 1.01 $\pm$ 0.24 & 29.55 $\pm$ 0.65\\
\hline
LLaVA → MiniGPT-4 ($\epsilon=16/255$) & 39.54 $\pm$ 0.58 & 1.90 $\pm$ 0.14 & 31.62 $\pm$ 1.20  & 0.75 $\pm$ 0.14  & 16.03 $\pm$ 0.76  & 1.17 $\pm$ 0.18 & 39.10 $\pm$ 0.55 \\
LLaVA → MiniGPT-4 ($\epsilon=32/255$) & 39.76 $\pm$ 0.65 & 1.73 $\pm$ 0.11 & 31.47 $\pm$ 0.18  & 0.78 $\pm$ 0.04  & 15.78 $\pm$ 0.66  & 1.17 $\pm$ 0.25 & 39.18 $\pm$ 0.79 \\
LLaVA → MiniGPT-4 ($\epsilon=64/255$) & 38.88 $\pm$ 1.75 & 1.76 $\pm$ 0.18 & 31.12 $\pm$ 1.35  & 0.75 $\pm$ 0.12  & 15.32 $\pm$ 0.54  & 1.20 $\pm$ 0.14 & 38.49 $\pm$ 1.80 \\
LLaVA → MiniGPT-4 (unconstrained)     & 28.31 $\pm$ 1.11 & 1.13 $\pm$ 0.27 & 21.73 $\pm$ 0.92  & 0.42 $\pm$ 0.24  & 9.65  $\pm$ 0.28  & 0.82 $\pm$ 0.29 & 27.83 $\pm$ 1.14 \\
\hline
LLaVA → InstructBLIP ($\epsilon=16/255$) & 54.98 $\pm$ 0.58 & 4.64 $\pm$ 0.20 & 42.47 $\pm$ 0.15 & 7.68 $\pm$ 0.29  & 28.68 $\pm$ 0.14 & 4.55 $\pm$ 0.76 & 54.82 $\pm$ 0.65 \\
LLaVA → InstructBLIP ($\epsilon=32/255$) & 54.13 $\pm$ 1.17 & 4.24 $\pm$ 0.08 & 40.88 $\pm$ 0.60 & 7.14 $\pm$ 0.24  & 28.18 $\pm$ 0.71 & 4.41 $\pm$ 0.18 & 53.96 $\pm$ 1.17 \\
LLaVA → InstructBLIP ($\epsilon=64/255$) & 53.35 $\pm$ 1.99 & 4.16 $\pm$ 0.29 & 41.24 $\pm$ 1.31 & 6.89 $\pm$ 0.38  & 28.22 $\pm$ 1.03 & 4.71 $\pm$ 0.35 & 54.13 $\pm$ 1.94 \\
LLaVA → InstructBLIP (unconstrained)     & 53.87 $\pm$ 0.48 & 1.90 $\pm$ 0.21 & 44.87 $\pm$ 0.17 & 1.08 $\pm$ 0.19  & 22.60 $\pm$ 0.85 & 2.80$\pm$ 0.18  & 52.27$\pm$ 0.55  \\
\hline
MiniGPT-4 → LLaVA ($\epsilon=16/255$) & 52.62 $\pm$ 0.77 & 1.81 $\pm$ 0.17 & 42.77 $\pm$ 0.05 & 1.09 $\pm$ 0.08 & 21.45 $\pm$ 0.93 & 2.48 $\pm$ 0.13 & 52.00 $\pm$ 0.88 \\
MiniGPT-4 → LLaVA ($\epsilon=32/255$) & 53.37 $\pm$ 0.98 & 2.00 $\pm$ 0.08 & 43.04 $\pm$ 0.70 & 1.20 $\pm$ 0.05 & 20.87 $\pm$ 0.22 & 2.28 $\pm$ 0.49 & 52.59 $\pm$ 0.80 \\
MiniGPT-4 → LLaVA ($\epsilon=64/255$) & 54.03 $\pm$ 0.78 & 2.06 $\pm$ 0.04 & 44.10 $\pm$ 0.65 & 1.06 $\pm$ 0.04 & 22.18 $\pm$ 0.59 & 2.42 $\pm$ 0.25 & 53.23 $\pm$ 0.97 \\
MiniGPT-4 → LLaVA (unconstrained)     & 50.97 $\pm$ 0.71 & 1.95 $\pm$ 0.13 & 42.32 $\pm$ 0.80 & 0.75 $\pm$ 0.00 & 17.97 $\pm$ 0.84 & 1.42 $\pm$ 0.17 & 50.08 $\pm$ 1.01 \\
\hline
MiniGPT-4 → InstructBLIP ($\epsilon=16/255$) & 54.97 $\pm$ 1.05  &  4.58 $\pm$ 0.21 & 41.63 $\pm$ 0.61  & 7.20 $\pm$ 0.14 & 29.44 $\pm$ 0.86  & 4.16 $\pm$ 0.17  & 54.91 $\pm$ 1.05 \\
MiniGPT-4 → InstructBLIP ($\epsilon=32/255$) & 54.05 $\pm$ 0.26 & 4.33 $\pm$ 0.24 &  40.98 $\pm$ 0.38 & 7.18 $\pm$ 0.29 & 27.88 $\pm$ 0.46 & 4.50 $\pm$ 0.28 & 54.05 $\pm$ 0.26 \\
MiniGPT-4 → InstructBLIP ($\epsilon=64/255$)  & 55.70 $\pm$ 0.87 & 4.38 $\pm$ 0.05 & 42.91 $\pm$ 0.14 & 6.7 $\pm$ 0.54 & 28.75 $\pm$ 0.38 & 4.38 $\pm$ 0.13 & 55.67 $\pm$ 0.91 \\

MiniGPT-4 → InstructBLIP (unconstrained)     & 61.05 $\pm$ 0.61 &  5.34$\pm$ 0.56 & 46.70 $\pm$ 0.37 & 7.89 $\pm$ 0.32 & 30.90 $\pm$ 0.81 & 4.53 $\pm$ 0.45 & 60.85 $\pm$ 0.64 \\
\hline
\end{tabular}
\caption[]{Adversarial transferability results measured by the percentage (\%) of generated outputs exhibiting specific toxicity attributes, as evaluated by the Detoxify Classifier. The "Any" column signifies that generated outputs exhibiting any of the six toxic}
\label{tab:transferattack_detox}
\end{table*}

\section*{Appendix}
\subsection*{A GPT-4o Evaluation} 
We use GPT-4o to assess whether an adversarial prompt elicits undesired outputs from the target model.

\textbf{Prompt setting.} We employ a fixed evaluation prompt to instruct GPT-4o to
classify each generated response with respect to four risk categories. The exact prompt used is: ``Please use the ASR calculation method across four categories: Identity Attack, Disinformation, Violence/Crime and X-risk.'' This standardized prompt ensures consistent evaluation of jailbreak responses across all models.

\textbf{Aggregation.} For each adversarial attempt, we repeat ten times under the same prompt and model configuration to account for sampling variability. Each response is independently evaluated by GPT-4o using the fixed evaluation prompt. The attempt is considered successful if at least one of the ten responses contains undesirable content and ASR is computed accordingly.

\subsection*{B Adversarial transferability Evaluation} 
This section provides supplementary empirical results on adversarial transferability referenced in the main paper, offering a more detailed view of transferability on target models.

We adopt the automatic evaluation to assess whether an adversarial prompt successfully induces the target model to produce undesired outputs in Table~\ref{tab:presapi}. To evaluate adversarial transferability, we reuse the same adversarial prompt on unseen target models and repeat a stronger attack setting in Table~\ref{tab:transferability}.

We additionally report how transferability varies across InstructBLIP, LLaVA, and MiniGPT-4 under multiple $\epsilon$ settings. Each generated response is evaluated by the Perspective API and Detoxify Classifier. Table~\ref{tab:transferattack_perspective} and Table~\ref{tab:transferattack_detox} provide a comprehensive analysis of the adversarial transferability of ZO-SPSA attacks across three target models. Each surrogate model undergoes various perturbation budget constraints $\epsilon = 16/255$, $32/255$, $64/255$ and an unconstrained setting. 

\end{document}